\documentstyle[prl,aps,floats,psfig]{revtex} 
\begin{document}
\input epsf
\title{A coherent double-quantum-dot electron pump} 
\author{T. H. Stoof and Yu. V. Nazarov} 
\address{Department of Applied Physics,\\
Delft University of Technology, Lorentzweg 1, 2628 CJ Delft,
The Netherlands}
\address{\mbox{ }}
\address{\parbox{15cm}{\rm \mbox{ }\mbox{ }\mbox{ } 
We have investigated the transport characteristics of an electron pump 
consisting of an asymmetric double quantum dot at zero bias voltage which 
is subject to electromagnetic radiation. Depending on the energies of the 
intermediate states in the pumping cycle, electrons can be transferred 
through the dots incoherently via sequential tunneling or coherently via 
co-tunneling. The dc transport through the system can be controlled by the 
frequency of the applied radiation. We concentrate on resonant tunneling
peaks in the pumping current that are interesting to observe experimentally. 
}}
\twocolumn
\maketitle
A quantum dot can be thought of as an artificial atom with adjustable 
parameters. It is of more than fundamental interest to study its 
properties under various circumstances, e.g. by transport experiments\cite{leorev}. 
By considering 
a double-quantum-dot system, the analogy with real atoms can be stretched to 
include artificial molecules. The analogue of the covalent bond is then 
formed by an electron which coherently tunnels back and forth between 
the two dots. By applying electromagnetic radiation with a frequency equal 
to the energy difference between the time-independent eigenstates of the 
double-dot system, an electron can undergo these so-called spatial Rabi 
oscillations even when the tunneling matrix element between the dots is 
small~\cite{sn2,stafford2}. Recently, several time-dependent 
transport measurements on quantum-dot systems have been 
reported~\cite{spectro}, most of them aimed to characterize the discrete
states in the dots. It has also been suggested to make devices 
from quantum dots. Examples of such applications are pumps that transfer electrons 
one by one by using time-dependent voltages to alternatingly raise and lower 
tunnel barrier heights~\cite{leo40}, or systems in which coupled 
quantum dots (or quantum wells) are used for quantum-scale information 
processing~\cite{qcomp}. Because the transport mechanism in the 
abovementioned pumps is determined by sequential tunneling, electrons are pumped 
incoherently. In this paper we describe an electron pump that can also 
transfer electrons coherently by means of a co-tunneling process. In this case,
the system  switches coherently between two states each time an electron is
pumped through the dots. The number of electrons in the double dot does not
change, however, resulting in less current noise. Alternatively, this process
can be seen as coherent transport through a double-quantum-dot qubit. 
 
\begin{figure}[t] 
\hspace*{1cm} 
\epsfxsize=5cm \epsfbox[0 400 450 805]{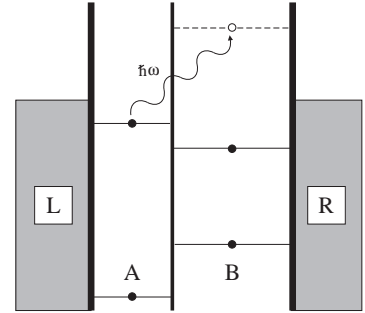} 
\caption{Asymmetric double-dot system at zero bias voltage. The dots A and B are 
coupled to the reservoirs L and R by tunnel junctions. An electron in the 
left dot can tunnel to the right by absorbing a photon.} 
\label{fig:ddotsys} 
\end{figure}

The system we consider consists of two weakly coupled quantum dots A and B 
connected to two large reservoirs L and R by tunnel barriers 
(see Fig.~\ref{fig:ddotsys}). 
The leads are assumed to have a continuous energy spectrum and the bias voltage 
is zero. 
Each individual quantum dot is in its ground state and in the remainder of this 
paper we will only concentrate on the case of transitions between ground states 
of the dots. 
Moreover, we assume that the double-dot system is asymmetric in the ground 
state $|0,0\rangle$, where $|n,m\rangle$ denotes a full many-body state 
with $n$ extra electrons in the left dot and $m$ in the right dot, 
respectively. This asymmetry entails that the energy of the state 
$|1,-1\rangle$ is much higher than the energy of the state $|-1,1\rangle$. 
In that case an electron can be excited from the left dot into the right 
one, but the probability of the reverse process occurring can be neglected. 
By fabricating two dots of different sizes and adjusting the energy levels with
a gate voltage, such an asymmetry can be easily realized. The energy difference 
between the ground state $|{ a}\rangle=|0,0\rangle$ and the excited state 
$|{ b}\rangle=|-1,1\rangle$ is denoted by $\varepsilon_{0}$. It can be tuned 
by an external gate voltage, and we assume it to be 
much larger than the transition matrix element between the dots $T$, i.e. 
$\varepsilon_{0} \gg T$. In this case dc transport through the 
double dot is blocked. 
 
This situation changes if we apply  
electromagnetic radiation to the system. Assume that a time-dependent 
oscillating signal is present on the gate electrode, so that the 
time-dependent energy difference between states $|{ a}\rangle$ and $|{ b}\rangle$ 
becomes $\varepsilon(t) = \varepsilon_{0} + \tilde{\varepsilon} \cos \omega t$, 
where $\tilde{\varepsilon}$ is the amplitude and $\omega$ the frequency 
of the externally applied signal. When the frequency of the applied 
radiation matches the time-independent energy difference between states 
$|{ a}\rangle$ and $|{ b}\rangle$, i.e. if $\omega \approx \varepsilon_{0}$, 
it is possible for an electron from the left dot to tunnel to the  
right one. In principle, this electron can now leave the system by 
tunneling to the right lead, resulting in state $|{ c}\rangle=|-1,0\rangle$. 
An electron from the left lead can then tunnel to the left dot, thus 
restoring the ground state. Effectively, an electron has now been 
transferred from the left electrode to the right one. This transport cycle, 
$|0,0\rangle \rightarrow |-1,1\rangle \rightarrow |-1,0\rangle  
\rightarrow |0,0\rangle$, is not the only one. Another possible sequence, 
in which the system passes the intermediate state $|{ d}\rangle=|0,1\rangle$, 
is given by $|0,0\rangle \rightarrow |-1,1\rangle \rightarrow |0,1\rangle  
\rightarrow |0,0\rangle$. 
 
\begin{figure}[h] 
\epsfxsize=6cm \epsfbox[65 585 375 800]{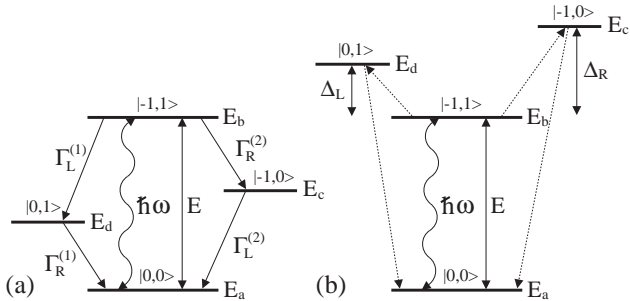} 
\caption{Energy diagrams of the double-dot system in which (a) only 
incoherent and (b) only coherent tunneling occurs. Indicated are the four 
possible states of the system and their respective energies: 
$|{  a}\rangle=|0,0\rangle$, $|{  b}\rangle=|-1,1\rangle$, 
$|{  c}\rangle=|-1,0\rangle$, and $|{  d}\rangle=|0,1\rangle$.} 
\label{fig:energy} 
\end{figure}

The details of the transport mechanism of a pumping cycle depend on 
the energies of states $|{ c}\rangle$ and $|{ d}\rangle$, 
as shown in Fig.~\ref{fig:energy}. 
If the energies of these intermediate states are somewhere between those 
of states $|{ a}\rangle$ and $|{ b}\rangle$, the double-dot system relaxes to
the ground state via the sequential (and thus incoherent) tunneling processes 
described above. As shown in Fig.~\ref{fig:energy}~(a), the four tunneling 
processes are described by the rates $\Gamma_{ L}^{(1)}$, 
$\Gamma_{ R}^{(1)}$, $\Gamma_{ L}^{(2)}$, and $\Gamma_{ R}^{(2)}$, 
respectively. Here, units are used such that 
$\hbar=1$. In the opposite case when the energies of the intermediate 
states are higher than that of state $|{ b}\rangle$,
Fig.~\ref{fig:energy}~(b), transport via 
sequential tunneling is blocked. However, transport is still possible 
via inelastic co-tunneling of electrons~\cite{cotunnel}. When the system 
is in state $|{ b}\rangle$, two electrons can tunnel simultaneously, 
one going from the left reservoir to the left dot and the other going from 
the right dot to the right electrode. In this process one of the intermediate 
states is virtually occupied. The necessary energy is provided by relaxing 
the system to the ground state $|a\rangle$, thereby releasing an energy $E$. 
In the following, we will describe these two different mechanisms 
quantitatively, starting with the sequential tunneling regime.  
 
We use the density-matrix approach developed in 
Ref.~\cite{dmatrix}. Disregarding tunneling to and from the leads, 
the Hamiltonian of the two-level system is given by: 
\begin{equation} 
\cal{H}= 
\left(\begin{array}{cc}  
-\frac{1}{2}(\epsilon_{0}+\tilde{\epsilon}\cos \omega t)& T\\  
T & \frac{1}{2}(\epsilon_{0}+\tilde{\epsilon}\cos \omega t) 
\end{array} \right), 
\label{hamiltonian2} 
\end{equation} 
where $T$ describes the coupling between the dots.  
We are interested in the particular case  where 
the energy level spacing is large, i.e. $\epsilon_{0} \gg T$. Rewriting 
Eq.~(\ref{hamiltonian2}) in the basis of eigenstates of the time-independent 
($\tilde{\epsilon}=0$) Hamiltonian and using the fact that 
$\tilde{\epsilon} \ll \epsilon_{0}$, we obtain: 
\begin{equation} 
\cal{H}= 
\left(\begin{array}{cc}  
-\frac{1}{2}E & \overline{T} \cos \omega t\\  
\overline{T} \cos \omega t & \frac{1}{2}E 
\end{array} \right), 
\label{newhamil} 
\end{equation} 
where $E=\sqrt{\epsilon_{0}^2+4 T^2}\approx \epsilon_{0}$ is the 
renormalized energy level spacing and 
$\overline{T}=T \tilde{\epsilon}/E$. From Eq.~(\ref{newhamil}) 
we see that there are small time-dependent matrix elements that cause 
mixing between the two states. Using the equation of motion for the density 
matrix, $i d\rho/dt=[\cal{H},\rho]$, including tunneling to and from the 
leads, and disregarding rapidly oscillating terms, we obtain the equations of 
motion for the density-matrix elements which are valid near resonance: 
\begin{mathletters} 
\begin{eqnarray} 
\label{een} 
\dot{\rho}_{aa} &=& 
+i\mbox{$\frac{1}{2}$}\overline{T} (\rho_{ab}-\rho_{ba}) 
+\Gamma_{ L}^{(2)}\rho_{cc} + \Gamma_{ R}^{(1)}\rho_{dd},\\ 
\label{twee} 
\dot{\rho}_{bb} &=& 
-i\mbox{$\frac{1}{2}$}\overline{T} (\rho_{ab}-\rho_{ba})-\Gamma_{ T}\rho_{bb},\\ 
\label{drie} 
\dot{\rho}_{cc} &=& 
\Gamma_{ R}^{(2)}\; \rho_{bb}-\Gamma_{ L}^{(2)}\; \rho_{cc}, \\ 
\label{vier} 
\dot{\rho}_{ab} &=& 
i\mbox{$\frac{1}{2}$}\overline{T}(\rho_{aa}-\rho_{bb}) + i(E-\omega)\rho_{ab} - 
\mbox{$\frac{1}{2}$}\Gamma_{ T} \rho_{ab}, 
\end{eqnarray} 
\end{mathletters} 
where $\Gamma_{ T}=\Gamma_{ L}^{(1)} + \Gamma_{ R}^{(2)}$ and $\rho_{aa}$, 
$\rho_{bb}$, $\rho_{cc}$, and $\rho_{dd}=1-\rho_{aa}-\rho_{bb}-\rho_{cc}$ 
denote the probabilities for an electron to be in the states $|{ a}\rangle$, 
$|{ b}\rangle$, $|{ c}\rangle$ and $|{ d}\rangle$, respectively. 
In the nondiagonal elements $\rho_{ab}=\rho_{ba}^{*}$, only the frequency parts 
$\rho_{ab}\exp(i\omega t)$ and $\rho_{ba}\exp(-i\omega t)$ that contribute to 
the resonant current peak have been retained. 
 
\begin{figure}[b] 
\vspace*{-1 cm} 
\hspace*{0.2 cm} 
\psfig{file=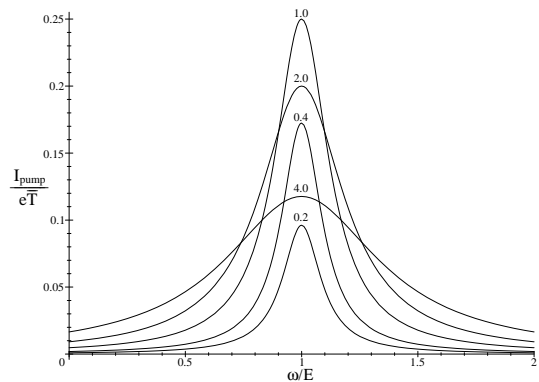,width=8.5cm,angle=270} 
\vspace*{-1 cm} 
\caption{The pumping current as a function of the applied frequency 
for different values of $\Gamma/\overline{T}$, where we have taken 
$\Gamma_{ L}^{(1)}=\Gamma_{ R}^{(1)}=\Gamma_{ L}^{(2)}=\Gamma_{ R}^{(2)} 
=\Gamma$. Calculated for $\tilde{\varepsilon}=0.1 \varepsilon_{0}$ and 
$T=0.1 \varepsilon_{0}$.} 
\label{fig:incopeak} 
\end{figure}

The average current through the system is given by: 
\begin{equation} 
\label{current} 
I/e=\Gamma_{ R}^{(2)}\rho_{bb}+\Gamma_{ R}^{(1)}\rho_{dd}. 
\end{equation} 
In Fig.~\ref{fig:incopeak} we have plotted a resonant current peak 
as a function of the applied radiation frequency for different values of
$\Gamma/\overline{T}$, where 
$\Gamma = \Gamma_{ L}^{(1)}= \Gamma_{ R}^{(1)}= \Gamma_{ L}^{(2)}= 
\Gamma_{ R}^{(2)}$.  
The current peak height initially grows with increasing $\Gamma$,  
its width remaining almost constant. The best resonance condition 
occurs for $\Gamma = \overline{T}$. When $\Gamma$ increases beyond this 
value, the height of the peak rapidly decreases while the width increases. 
Using an approach developed in Ref.~\cite{sn2}, which is valid near 
resonance, we are able to derive an analytical expression for the shape of 
the current peak in this regime. We find that the peak is a Lorentzian: 
\begin{mathletters} 
\begin{eqnarray} 
\label{lorentzian} 
I/e&=&I_{ max} w^2/[w^2+4(\omega-E)^2],\\ 
\label{height2} 
I_{ max}&=&\overline{T}^{2} \Gamma_{ T}/ 
[\Gamma_{ T}^2+(2+\alpha)\overline{T}^{2}],\\ 
\label{width2} 
w&=&\sqrt{\Gamma_{ T}^2+(2+\alpha)\overline{T}^2}, 
\end{eqnarray} 
\end{mathletters} 
where $\alpha=\Gamma_{ L}^{(1)}/\Gamma_{ R}^{(1)}+ 
\Gamma_{ R}^{(2)}/\Gamma_{ L}^{(2)}$. 
In the limit of small $\overline{T}$, i.e. in the limit of small radiation 
amplitude $\tilde{\varepsilon}$ and small overlap $T$, the height of the 
current peak is proportional to $\overline{T}^{2}$; $I_{ max}= 
\overline{T}^{2}/\Gamma_{ T}$, whereas its width 
is constant in this limit; $w=\Gamma_{ T}$. 
This concludes the discussion of the incoherent tunneling regime. 
 
In the coherent mechanism, two electrons tunnel simultaneously; one going 
from the left lead to the left dot, and one from the right dot to the right lead. 
Because the transport occurs via the virtual occupation of a state with a large 
electrostatic energy, these two tunneling events cannot be treated 
independently\cite{cotunnel}. Using Fermi's Golden Rule we obtain the  
zero-temperature co-tunnel rate\cite{thesis}: 
\begin{eqnarray} 
\nonumber 
\Gamma_{ ct}&=&\frac{2C\left[\ln(1+E/\Delta_{ L})+ 
\ln(1+E/\Delta_{ R})\right]}{\Delta_{ L}+ \Delta_{ R}+E}+\\ 
&&\frac{C E}{\Delta_{ L} \left(\Delta_{ L}+E \right)} + 
\frac{C E}{\Delta_{ R} \left(\Delta_{ R}+E \right)}, 
\label{ztctr} 
\end{eqnarray} 
where $C=2\pi | t_{ L} t_{ R}|^{2} \nu_{ L} \nu_{ R}$, and where 
the matrix elements for tunneling trough the left and right barier, 
$t_{ L,R}$, and the density of states in the left and right electrode, 
$\nu_{ L,R}$, are assumed to be energy independent. 
In the limit of large charging energies, $\Delta_{ L}, \Delta_{ R} \gg E$, 
which we will consider from now on, Eq.~(\ref{ztctr}) reduces to a simple 
expression: 
\begin{equation} 
\Gamma_{ ct} = 2\pi | t_{ L} t_{ R}|^{2} \nu_{ L} \nu_{ R}  
\left( \frac{1}{\Delta_{ L}} + \frac{1}{\Delta_{ R}} \right)^{2} E. 
\label{simple2} 
\end{equation} 
The prefactor $C$ is proportional to the product of the individual tunnel rates
through the left and right barrier, and the energy denominators reflect the fact 
that the tunneling occurs via the virtual occupation of two states. Because
co-tunneling is a second-order process, this co-tunnel rate will be much
smaller than the sequential tunnel rates of the incoherent regime.
In the absence of radiation, the average current through the system is given by: 
\begin{equation} 
I/e = \Gamma_{ ct} \rho_{bb}, 
\label{cohercur} 
\end{equation} 
which is zero, since $\rho_{bb}=0$. 
In order to calculate the current in the presence of radiation, we therefore 
need equations for the density matrix elements $\rho_{aa}$ and $\rho_{bb}$ 
taking into account the radiation-induced co-tunneling processes. 
In contrast to the situation in the incoherent mechanism, these density 
matrix elements are the only diagonal ones because the states 
$|{ c}\rangle$ and $|{ d}\rangle$ are now occupied only virtually. 
 
As before, we choose the eigenstates of the time-independent Hamiltonian as 
the basis for our calculations. The time-dependent Schr\"odinger equation 
then becomes: 
\begin{equation} 
i d \vec{\psi}_{\alpha}(t)/d t ={\cal H}(t) \vec{\psi}_{\alpha}(t), 
\label{schroedeq} 
\end{equation} 
where ${\cal H}$ is given by Eq.~(\ref{newhamil}) and $\alpha= a,b$. 
Expanding the wave functions $\psi_{\alpha}(t)$ in harmonics we obtain: 
\begin{equation} 
\vec{\psi}_{\alpha}(t) = e^{- i E_{\alpha} t} \sum_{n} \vec{c}^{n}_{\alpha} 
e^{-i n \omega t}. 
\label{expansion} 
\end{equation} 
In the case of a small radiation amplitude, 
$\tilde{\varepsilon} \ll \varepsilon_{0}$, and near resonance, 
$\omega \approx E$, it is a good approximation to retain only the $n=0$ and 
$n=1$ terms for $\vec{\psi}_{ a}(t)$, and the $n=0$ and $n=-1$ terms for 
$\vec{\psi}_{ b}(t)$. This approximation entails that we only 
take into account single-photon transitions between the states $|{ a}\rangle$ 
and $|{ b}\rangle$. The wave functions then become: 
\begin{mathletters} 
\begin{eqnarray} 
\label{approxa} 
\vec{\psi}_{ a}(t)&=&e^{-i E_{ a} t}~( \xi \vec{\psi_{ a}} - 
\zeta e^{-i \omega t} \vec{\psi_{ b}} ),\\ 
\label{approxb} 
\vec{\psi}_{ b}(t)&=&e^{i E_{ a} t}~( \zeta e^{i \omega t} 
\vec{\psi_{ a}}+ \xi \vec{\psi_{ b}} ), 
\end{eqnarray} 
\end{mathletters} 
where the coefficients $\xi$ and $\zeta$ are given by: 
\begin{mathletters} 
\begin{eqnarray} 
\label{xi} 
\xi&=&(\delta E+W)/\sqrt{2W(W+\delta E)},\\ 
\label{zeta} 
\zeta&=&\overline{T}/\sqrt{2W(W+\delta E)}, 
\end{eqnarray} 
\end{mathletters} 
where $\delta E=E-\omega$ and $W^2={\delta E}^{2}+\overline{T}^{2}$. 
At resonance, when $\omega = E$, the coefficients are $\xi=\zeta= 
\frac{1}{2}\sqrt{2}$. The energy $E_{ a}$ in Eqs.~(\ref{approxa}) and 
(\ref{approxb}) is given by: 
\begin{equation} 
E_{ a}=-E_{ b}=-\mbox{$\frac{1}{2}$}E-\mbox{$\frac{1}{2}$}W, 
\label{enul} 
\end{equation} 
reflecting the fact that the energy levels $-\frac{1}{2}E$ and $\frac{1}{2}E-\omega$ 
slightly repel each other when $\omega \approx E$.
 
With the aid of the closed-time-path Green function technique~\cite{pathint},
which was applied to quantum-dot systems in Ref.~\cite{herbert}, the wave
functions Eqs.~(\ref{approxa}) and (\ref{approxb}) are used to obtain the
equations of motion for the density-matrix elements near resonance. Without
going into the details of this calculation (they can be found in
Ref.~\cite{thesis}), we simply present here the resulting set of equations: 
\begin{mathletters} 
\begin{eqnarray} 
\label{tunratefreq} 
\dot{\rho}_{bb}&=&-i\mbox{$\frac{1}{2}$}\overline T (\rho_{ab}-\rho_{ba})- 
\Gamma_{ ct} \rho_{bb} + \xi^{2} \zeta^{2} \Gamma_{ ct},\\ 
\dot{\rho}_{ab}&=&i\mbox{$\frac{1}{2}$} \overline{T} (\rho_{aa}-\rho_{bb})+ 
i(E-\omega)\rho_{ab}-\mbox{$\frac{1}{2}$} \Gamma_{ ct} \rho_{ab}, 
\end{eqnarray} 
\end{mathletters} 
where $\rho_{aa}+\rho_{bb}=1$, $\rho_{ba}=\rho_{ab}^{*}$, and where we have used 
$\omega \approx E$ and 
$E_{ a} \approx -E/2$ to calculate the dissipative terms. Clearly these 
equations are very similar to Eqs.~(\ref{twee}) and (\ref{vier}). The crucial 
difference with the equations for the density matrix in the incoherent mechanism, 
however, is that the extra tunnel rate in Eq.~(\ref{tunratefreq}) depends on 
the applied frequency and radiation amplitude via $\xi$ and $\zeta$: It is a 
Lorentzian centered around $\omega = E$ having width $2 \overline{T}$. 
 
\begin{figure}[h] 
\vspace*{-1cm} 
\hspace*{0.2cm} 
\psfig{file=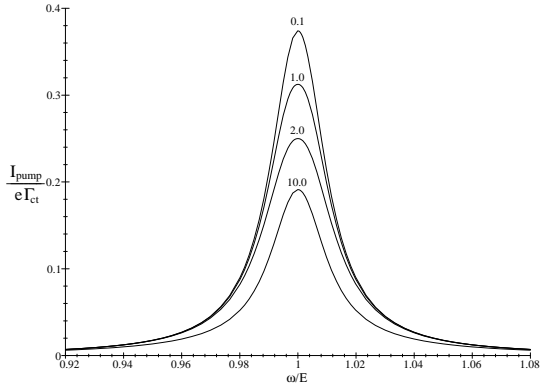,width=8.5cm,angle=270} 
\vspace*{-1cm} 
\caption{The scaled pumping current in the coherent regime as a function 
of the applied frequency for different values of $\Gamma_{ ct}/\overline{T}$. 
Calculated for $\tilde{\varepsilon}=0.1 \varepsilon_{0}$ and $T=0.1 \varepsilon_{0}$.} 
\label{fig:cotunpeak} 
\end{figure}

We solve for the stationary solution of these equations to calculate the average 
current through the system which is given by the probability $\rho_{bb}$ to be
in the excited state $|b\rangle$ times the decay rate of that state: 
\begin{equation} 
I/e=(\xi^{4}+\xi^{2} \zeta^{2}+\zeta^{4}) 
\Gamma_{ ct} \rho_{bb}. 
\label{endcur} 
\end{equation}  
In Fig.~\ref{fig:cotunpeak} the scaled pumping current is plotted for different 
values of $\Gamma_{ ct}/\overline{T}$. 
With increasing co-tunnel rate, the scaled peak height first drops rapidly and 
then decreases slowly to its minimum value, while its width changes only slightly. 
The peak shape is a Lorentzian of the form Eq.~(\ref{lorentzian}) with height 
$I_{ max}$ and width $w$: 
\begin{mathletters} 
\begin{eqnarray} 
\label{heightcot} 
I_{ max}&=&\mbox{$\frac{3}{16}\Gamma_{ ct}$}  
(4 \overline{T}^2+\Gamma_{ ct}^2)/(2 \overline{T}^2+ \Gamma_{ ct}^2),\\ 
\label{widthcot} 
w&=&\overline{T} \sqrt{6(2\overline{T}^2+\Gamma_{ ct}^2) 
(4\overline{T}^2+\Gamma_{ ct}^2)/(8\overline{T}^4+\Gamma_{ ct}^4}). 
\end{eqnarray} 
\end{mathletters} 
The height $I_{ max}$ reduces to $\frac{3}{8}\Gamma_{ ct}$ for small 
$\Gamma_{ ct}$ and to $\frac{3}{16} \Gamma_{ ct}$ for large $\Gamma_{ ct}$.  
A comparison with Fig.~\ref{fig:incopeak} clearly shows that the co-tunnel current peak 
is much narrower than the incoherent one. Another distinct feature of this  
co-tunnel peak is the fact that its width changes nonmonotonically with 
increasing co-tunnel rate $\Gamma_{ ct}$. For small values of 
$\Gamma_{ ct}$ the width is equal to $w = \sqrt{6}~\overline{T} \approx 
2.45~\overline{T}$. It then increases rapidly to reach a maximum of 
$w = \sqrt{6+\frac{9}{2}\sqrt{2}}~\overline{T} \approx 3.52~\overline{T}$ for 
$\Gamma_{ ct}= 2^{3/4}~\overline{T} \approx 1.68~\overline{T}$. On increasing 
the co-tunnel rate further, the width decreases again to $\sqrt{6}~\overline{T}$. 
The width is thus comparatively insensitive to changes in the co-tunnel rate. 
This is due to the fact that the extra co-tunnel rate in Eq.~(\ref{tunratefreq})
itself has an intrinsic width $w=2 \overline{T}$, thus restricting the width
of the current peak dramatically. 
 
In conclusion, we considered an electron pump consisting of a  
double quantum dot subject to radiation. An incoherent and a coherent 
pumping mechanism has been discussed. By deriving equations of motion for the 
density matrix elements of the double-dot system, we calculated the 
pumping current in both regimes. Whereas the tunnel rates as a function of the 
external frequency are constants in the incoherent mechanism, they are 
Lorentzians in the co-tunneling regime. In both cases the current peak is a 
Lorentzian, but in the coherent case the peak width is much smaller than in 
the sequential-tunneling regime and changes nonmonotonically as a function 
of the co-tunnel rate. Experimental realization of this device would allow 
for a systematic study of coherent transport through a solid-state qubit.  
 
It is a pleasure to acknowledge useful discussions with Gerrit Bauer, 
Henk Stoof, Caspar van der Wal, Tjerk Oosterkamp, and Leo Kouwenhoven. 
This work is part of the research program of the "Stichting voor 
Fundamenteel Onderzoek der Materie" (FOM), which is financially 
supported by the "Nederlandse Organisatie voor Wetenschappelijk  
Onderzoek" (NWO).

\end{document}